\documentclass[12pt]{article}
\begin{document}
\title{Motion of a thin spherically symmetric Shell of  Dust in the Schwarzschild field}
\author{Hans-J\"urgen  Schmidt}
\date{September  10, 2014}
\maketitle
\centerline{Institut f\"ur Mathematik, Universit\"at Potsdam, Germany} 
\centerline{Am Neuen Palais 10, D-14469 Potsdam,  \  hjschmi@rz.uni-potsdam.de}
\begin{abstract}
Summary. The equation of motion announced in the title was already 
deduced for the cases the inner metric being flat and the shell being 
negligibly small (test matter), using surface layers and geodesic trajectories resp. 
Here we derive the general equation of motion and solve it in closed form for
the case of parabolic motion. Especially the motion near the horizon and near the
singularity are examined. \par
Reprinted from: 10th International Conference on General Relativity and 
Gravitation, Padova (Italy) July 4 - 9, 1983. Eds.: B. Bertotti, 
F. de Felice, A. Pascolini, Contributed papers Vol. 1, 
Consiglio Nazionale Delle Ricerche, Roma (1983) page 339 - 341. \par
Author's address at that time: Zentralinstitut f. Astrophysik d.
Akademie d. Wissenschaften d. DDR -- 1502 Potsdam-Babelsberg
\end{abstract}

\section{} 
 Motion represents a problem in General Relativity even in such cases where the
particle's interaction is gravitational only. Because exact solutions of the problem are
very rare many approximate solutions have been calculated, but they
are often unsatisfactory for various reasons. On the other  hand, surface layers
 are exact solutions of  Einstein's field  equations, if these equations are written as 
distribution  equations. Furthermore, considering spherically symmetric
 configurations the problem of gravitational radiation does not  arise. Therefore  it is 
appropriate to consider the general motion of a spherically symmetric surface layer. 

\section{} 
Surface layers can be thought as strata of matter with a  negligible small thickness; 
they can be represented by  (non-spurios)  jumps of the Christoffel affinities -- their 
derivatives  entering the energy-momentum tensor yield $\delta $-distributions.
The related formulas are deduced in [1], [2] and [3]. Another but equivalent approach 
[4] uses the second fundamental tensor   and will be applied here to deduce the above 
mentioned motion. The cases where the interior metric is flat and the matter 
 is dust are treated in [4]. In [5], additionally, charges have been considered. 
 A static configuration is discussed in [7], and the tangential pressure 
is due to collisions of particles moving on circular orbits there. 

 \section{} 
Now let $\Sigma $ be the spherically symmetric time-like hypersurface
 at which matter is concentrated. Inside $\Sigma $ we take proper time 
$T$ and usual angles $\psi$, $\varphi$ as coordinates, 
$d \Omega^2 = d\psi^2 + \sin^2 \psi d \varphi ^2$. The invariant surface of 
 the sphere $T=$ const. will be denoted by $4 \pi R^2$, $R=R(T)$.
Then the first fundamental tensor of $\Sigma $ is $g_{\alpha \beta}$, and
$$
ds^2 = g_{\alpha \beta} d\xi^{\alpha} d\xi^{\beta} =
  - dT^2 + R^2 d \Omega^2, \, \xi^0 = T, \, \xi^2 = \psi, \, \xi^3 = \varphi.
$$
Outside $\Sigma $ one has to use the Schwarzschild metric with mass
parameters $M_-$ and $M_+$ in $V_- = [r < R]$ and $V_+ = [r > R]$
resp. and to match them together in a continuous manner.  \par
 Energy conditions require $M_+ \ge M_- \ge 0$, $M_-$ is the black
hole's mass and $M_+ - M_-$ is the shell's gravitational mass.  \par
The  energy momentum tensor is $(\xi^1 = r)$, $(i, \, k = 0, \, 1, \, 2, \, 3)$
$$
T_i^k = \delta(r-R) \left[ {\rm diag} (- \mu, \, 0, \, p, \, p) \right]_i^k
$$
with ($G=c=1$) 
$$
\mu = \frac{W_- - W_+}{4\pi R}, \qquad W_{\pm}
= \sqrt{1 + (dR/dT)^2 - 2M_{\pm} / R}
$$
and
$$
8\pi p = \left[  d^2R/dT^2  + (1 + (dR/dT)^2)/R \right] \left( W_+^{-1} - W_-^{-1}
 \right) + \frac{M_-}{R^2 W_-} -  \frac{M_+}{R^2 W_+} 
$$
(cf. [8] for details). There is no $T_1^k$ -component because by construction
no matter flow orthogonal to $\Sigma $ exists, and the tangential pressure is 
isotropic because of spherical symmetry. \par
 Requiring an equation of state $p = \alpha \mu$ one obtains a differential equation 
for the function $R(T)$: 
$$
R  d^2R/dT^2  = 2 \alpha W_+ W_- - 
\frac{M_+W_- +  M_- W_+}{R(W_+ + W_-)}\, .
$$
The conservation law $T^i_{0;i} = 0$ leads to 
$$
\mu  \cdot 4 \pi R^2 R^{2\alpha} = E = {\rm const.},
$$
and therefore 
$$
R (dR/dT)^2 = M_+ + M_- + \frac{E^2}{4 R^{4\alpha + 1}}
+ (M_+ - M_-)^2 R^{4\alpha + 1} / E^2 - R \,  . 
$$
For $\alpha =0$, there of course $E = M_+ - M_-$, we have the case
of matter being dust. For $p \le 0$, the shell will collapse into the
singularity after a finite proper time interval.  \par
Parabolic motion takes place if $dR/dT \to 0$ as $R \to \infty$ and is therefore
 possible for $\alpha = 0$ only. For this case the solution, which describes the 
fall into $R=0$ at $T=0$, is
$$
R(T) = \frac{H \cdot \sqrt[3]{(M_+ + M_-)/2}
}{ \sqrt[3]{-3T - K + H} - \sqrt[3]{-3T - K - H} }
$$
where 
$$
H = \sqrt{9 T^2 + 6 KT} \qquad {\rm and} \qquad
2K = (M_+ - M_-  )^3 / ( M_+ + M_-)^2 \, . 
$$

 \section{} 
The horizon is not a singular surface (but its geometry has an interesting 
shape, cf. [8]) and therefore its crossing cannot be measured locally. 
This statement keeps valid if  {\it locally} means {\it all information from a 
neighbourhood}, e.g. all curvature.

 \section{} 
The limiting process from a gravitational field creating particle to a test particle 
leads to geodesic motion (cf. e.g. [6]), but in the proof there  is presumed the 
curvature to be bounded in a neighbourhood  of the particle's world line. The question 
arises whether this is an essential  presumption or not.  Here we have a similar 
configuration: By a formal inserting of $\alpha = E = 0$ one obtains the equation 
$$
R d^2R/dT^2  = - M_+ / R, 
$$
which is indeed the equation of a radially falling geodesic test particle. 
But in the limiting process $ E \to 0$, $R=0$ is a singular point of the 
differential equation, and indeed, the leading  terms  in a  neighbourhood 
of $T=0$ are
$$
R(T) \sim T^{1/2} \qquad {\rm for \ each} \qquad  M_+ > M_-
$$
and
$$
R(T) \sim T^{2/3} \qquad {\rm for \ } \qquad  \qquad  M_+ = M_- \, .
$$
The typical radius where the test particle's motion and the layer's motion 
 begin to differ is 
$$
R(T) \approx \frac{(M_+ - M_-)^2}{M_+} \, .
$$
This means the dust strata  approach the singularity with another power of
proper time than the related geodesic test particle does. This supports the 
common opinion on this question by a sequence of exact solutions.

\noindent 
Note added in 2014: Reference  [8] appeared as
H.-J. Schmidt: Surface layers in General Relativity and their
relation to surface tensions, Gen. Rel. Grav.  {\bf  16} (1984) 
1053 - 1061; see arXiv:gr-qc/0105106v1  for a reprint. 

\end{document}